\documentstyle[12pt,amstex,amssymb,amscd,epsfig]{article}
\def\Order#1{${\cal O}(#1$)}

\reversemarginpar
\allowdisplaybreaks
\begin{document}
 
\begin{titlepage}
 
\begin{flushright}
{\bf 
     UTHEP-95-1001}
\end{flushright}

\vspace{0.1cm}
\begin{center}
{\LARGE
BHWIDE 1.00: \Order{\alpha} YFS Exponentiated Monte\\
Carlo for Bhabha Scattering at Wide Angles\\
for LEP1/SLC and LEP2$^{\dagger}$
}
\end{center}
 
\begin{center}
 {\bf S. Jadach}\\
   {\em Institute of Nuclear Physics,
        ul. Kawiory 26a, Krak\'ow, Poland}\\
   {\em CERN, Theory Division, CH-1211 Geneva 23, Switzerland,}\\
 {\bf W.  P\l{a}czek$^{\star}$}\\
   {\em Department of Physics and Astronomy,\\
   The University of Tennessee, Knoxville, Tennessee 37996-1200}\\
 {\bf B.F.L. Ward}\\
   {\em Department of Physics and Astronomy,\\
   The University of Tennessee, Knoxville, Tennessee 37996-1200\\
   SLAC, Stanford University, Stanford, California 94309}\\
   and \\
   {\em CERN, Theory Division, CH-1211 Geneva 23, Switzerland}
\end{center}

\vspace{1.0cm}
\begin{abstract}
We present \Order{\alpha} YFS exponentiated results for wide
angle Bhabha scattering at LEP/SLC energies using a new Monte
Carlo event generator BHWIDE~1.xx. Our calculations include
two options for the pure weak corrections, as presented in 
Beenakker \underline{et al.} and in B\"ohm \underline{et al.} From 
comparison with  the results of Beenakker \underline{et al.},
Montagna \underline{et al.} and Cacciari \underline{et al.}, we conclude
that the total precision of our BHWIDE results is $0.3\%(0.5\%)$ in the
LEP1/SLC regime within $\pm 100$ MeV ($+2.75/-2.5$ GeV) of the Z peak.
For LEP2, the corresponding precision is 
currently estimated at $1.5\%$; the latter could be improved if the data
in LEP2 so require.  Both precision tags represent clear improvements
over what is currently available in the literature. 
\end{abstract}
 
\begin{center}
{\it Submitted to Physics Letters}
\end{center}
 
\vspace{0.5cm}
\renewcommand{\baselinestretch}{0.1}
\footnoterule
\noindent
{\footnotesize
\begin{itemize}
\item[${\dagger}$]
Work supported in part by the US DoE contract DE-FG05-91ER40627,
Polish Government grants KBN 2P30225206, 2P03B17210 
and Polish-French Collaboration
within IN2P3.
\item[${}^{\star}$]
On leave of absence from Institute of Computer Science, Jagellonian
University, Krak\'ow, Poland.
\end{itemize}
}
 
\begin{flushleft}
{\bf 
     UTHEP-95-1001\\
     Oct.  1995 }
\end{flushleft}
 
\end{titlepage}
 
%===================INTRODUCTION============================================
\section{Introduction} 
As the final LEP1 data analysis begins and the initial stages of LEP2
materialize while the SLD prepares for the beginning of what may be its
final phase, the subject of precision calculations of wide angle Bhabha
scattering becomes more and more interesting, particularly from the 
standpoint of realizing such calculations via a Monte Carlo event generator
which would allow comparison between theory and experiment at the level
of events in the presence of {\it arbitrary} detector cuts. Indeed, currently,
the work of Beenakker \underline{et al.}~\cite{alibaba:1991a} 
and of Montagna \underline{et al.}~\cite{montagna:1993} 
represent the state of the art in the 
theoretical arena of the prespective calculations, and neither of these
approaches has provided a genuine Monte Carlo event generator to
allow comparison with arbitrarily cut experimental data. This has had the 
effect that the important $Z$ physics parameter $\Gamma_{e\bar e}$
has been extracted from the data using a subtraction of the
contamination from the t-channel $\gamma$-exchange
in the wide angle acceptance relevant
to its measurement. A Monte Carlo event generator with sufficient precision
would obviate the need for this procedure, for example. Motivated by
this and other such $Z$ and LEP2 
applications, we have developed an ${\cal O}(\alpha)$
Yennie-Frautschi-Suura (YFS) \cite{yfs:1961} exponentiated
Monte Carlo event generator for 
wide angle Bhabha scattering at LEP1/SLC and LEP2 energies. We present
this development in what follows.\par

Specifically, our starting point will be the \Order\alpha\ YFS exponentiated
Monte Carlo (MC) event generator BHLUMI~1.xx developed by two of us (S.J. and
B.F.L.W.) in ref.~\cite{bhlumi1:1989} for low angle Bhabha scattering
in the SLC/LEP luminosity regime $17\,mrad< {\theta}_{e,\bar e}< 150\,mrad$,
where $\theta_{e, \bar e}$ are the respective $e,\bar e$ CMS scattering
angles. In order to arrive at an event generator valid for wide angles
at both LEP1/SLC and at LEP2 to sufficient accuracy,
we have had to introduce the effects of the $Z$ exchange graphs
into the calculations presented in ref.~\cite{bhlumi1:1989} and we have
had to introduce the effects of the pure weak one-loop corrections
into those calculations as well. We stress that we have made
contact with the pioneering \Order\alpha\ MC program 
BABAMC of refs.~\cite{bohm:1988,babamc:1988}
as well as with the semi-analytical program ALIBABA of 
ref.~\cite{alibaba:1991a}
in that our pure weak corrections libraries are taken from these two cases:
the user of our new wide angle Bhabha MC, BHWIDE~1.xx, may choose which
library he uses and we shall discuss both of them in what follows.
Our exact hard bremsstrahlung amplitude, in the presence of the
full set of s-channel and t-channel $\gamma$ and $Z$ exchanges,
we shall compute explicitly in what follows, using methods
of the CALKUL-type~\cite{calkul:1982}, as formulated by\footnote{ 
                    We note that Kleiss and Stirling~\cite{kleiss:1985}
                    have introduced analogous methods as well.}
Xu \underline{et al.}~\cite{xu:1987}.
It is in this way that we have arrived at our new \Order\alpha\ YFS
exponentiated MC event generator BHWIDE~1.xx which simulates
realistic multiple photon radiative effects
for wide angle Bhabha scattering in the LEP1/SLC and
LEP2 energy regimes.
\par
We have compared our results with those of BABAMC, of the 
Monte Carlo integrator program SABSPV
of ref.~\cite{sabspv:1995}, of the semi-analytical
program TOPAZ0 of refs.~\cite{topaz0:1993,topaz0:1995}, 
of ALIBABA, of the MC event generator BHAGENE3 of 
refs.~\cite{jfield:1994,bhagene3:1995}, 
of the MC event generator BHAGEN95 of 
refs.~\cite{bhagen:1992,bhagen:1993,bhagen:1994,bhagen:1995},
and of the MC event generator UNIBAB of ref.~\cite{unibab:1994}.
On this basis, we have arrived at a reliable estimate of the total
precision of our results. The basis of this estimate is presented in detail
in ref.~\cite{bhabha-lepws:1995} and we will review only its main features
and the main features of the aforementioned attendant comparisons in what
follows. 
\par
More precisely, our BHLUMI~1.xx Monte Carlo event generator realizes
the process 
\begin{equation}
e^+(p_1) + e^-(q_1) \longrightarrow e^+(p_2) + e^-(q_2) \;
+ \gamma_1(k_1) + \ldots + \gamma_n(k_n)
\label{process} 
\end{equation}
via the YFS exponentiated cross section formula
\begin{equation}
d\sigma= e^{2\alpha\,Re\,B+2\alpha\,
\tilde B}\sum_{n=0}^\infty{1\over n!}\int\prod_{j=1}^n{d^3k_j\over k_j^0
}\int{d^4y\over(2\pi)^4}e^{iy(p_1+q_1-p_2-q_2-\sum_jk_j)+D}
\nonumber\cr
\qquad\bar\beta_n(k_1,\dots,k_n){d^3p_2d^3q_2\over p_2^0q_2^0},
\label{eqone}
\end{equation}
where the
real infrared function $\tilde B$ and the virtual infrared function $B$ are
given in 
refs.~\cite{yfs:1961,bhlumi1:1989,sjbw:1988,bhlumi2:1992,bennie:1987} 
(for definiteness,
we record them in the form we shall use presently, as this representation
is not readily available in the current literature and is essential
for practical applications of the type we pursue here),
and where we note the usual connections
\begin{align}
2\alpha\,\tilde B & = \int^{k\le K_{max}}{d^3k\over k_0}\tilde S(k),
\notag\\
D & =\int d^3k{\tilde S(k)\over k^0}
     \left(e^{-iy\cdot k}-\theta(K_{max}-k)\right)
\label{eqtwo}
\end{align} 
for the standard YFS infrared emission factor
\begin{equation}
\tilde S(k)= {\alpha\over4\pi^2}\left[Q_fQ_{f'}
%Q_{\overset{\tiny(-)}{f'}}
\left({p_1\over p_1\cdot k}-{q_1
\over q_1\cdot k}\right)^2+\dots\right],
\label{eqthree}
\end{equation} 
if $Q_f$ is the electric charge of $f$ in units of the positron charge. 
Here, the ``$\ldots$'' 
represent the remaining terms in $\tilde S(k)$ obtained from the one
given by respective substitutions of $Q_f$, $p_1$, 
%$Q_{\overset{\tiny(-)}{f'}}$
$Q_{f'}$, $   q_1$ with corresponding values for the
other pairs of the respective external charged legs according to the YFS
prescription in ref.~\cite{yfs:1961} (wherein due attention is taken to obtain
the correct relative sign of each of the terms in $\tilde S(k)$ according to
this latter prescription) and in refs.~\cite{yfs2:1990,yfs3-unpub}, $f\ne e$,
%${\overset{\tiny(-)}{f'}} 
$ f' = \bar f$. We have explicitly
the representations
\begin{align}
2\alpha\Re B(p_1,q_1,p_2,q_2) + 2\alpha\tilde{B}(p_1,q_1,p_2,q_2;k_m)
 =  R_1(p_1,q_1;k_m) + R_1(p_2,q_2;k_m) &
\notag\\
+ R_2(p_1,p_2;k_m) + R_2(q_1,q_2;k_m) - R_2(p_1,q_2;k_m) - R_2(q_1,p_2;k_m) &
\label{YFSff}
\end{align}
with
\begin{align}
R_1(p,q;k_m) & = R_2(p,q;k_m) + \left(\frac{\alpha}{\pi}\right) 
                                \frac{\pi^2}{2},
\label{R1}\\
R_2(p,q;k_m) & = \frac{\alpha}{\pi} \Bigg\{ \bigg(\ln\frac{2pq}{m_e^2} -1
                                            \bigg)
                 \ln\frac{k_m^2}{p^0q^0} + \frac{1}{2}\ln\frac{2pq}{m_e^2}
                -\frac{1}{2}\ln^2\frac{p^0}{q^0} 
               -\frac{1}{4}\ln^2\frac{(\Delta+\delta)^2}{4p^0q^0} 
\notag\\
              &  -\frac{1}{4}\ln^2\frac{(\Delta-\delta)^2}{4p^0q^0} 
                -\Re Li_2\left(\frac{\Delta+\omega}{\Delta+\delta}\right)
                -\Re Li_2\left(\frac{\Delta+\omega}{\Delta-\delta}\right)
\notag\\
              & -\Re Li_2\left(\frac{\Delta-\omega}{\Delta+\delta}\right)
                -\Re Li_2\left(\frac{\Delta-\omega}{\Delta-\delta}\right)
                +\frac{\pi^2}{3} -1 \Bigg\},
\label{R2}
\end{align}
where $\Delta=\sqrt{2pq+(p^0-q^0)^2}$, $\omega=p^0+q^0$, $\delta=p^0-q^0$,
and $k_m$ is a soft photon cut-off in the CMS 
($E_{\gamma}^{soft}<k_m\ll E_{beam}$).

\par
The YFS hard photon residuals $\bar\beta_i$ in~(\ref{eqone}), $i=0,1$,
are given in ref.~\cite{bhlumi1:1989} for BHLUMI~1.xx so that this latter
event generator calculates the YFS exponentiated exact
${\cal O}(\alpha)$ cross section for $e^+e^-\rightarrow e\bar e+ n(\gamma)$
with multiple initial, initial-final, and final state radiation 
using a corresponding Monte Carlo realization
of (\ref{eqone}) in the low angle regime of the SLC/LEP luminosity
monitor acceptance, where the $e^{\pm}$ scattering angles $\theta_{e^\pm}$
satisfy $17\,mrad< \theta_{e^{\pm}} < 150\,mrad$.  
In the next sections, we use explicit
Feynman diagrammatic methods and the results 
in refs.~\cite{alibaba:1991a,bohm:1988,babamc:1988}
to develop the corresponding Monte Carlo realization of
the respective application of~(\ref{eqone}) to the wide angle 
scattering regime, $\theta_{e^\pm}\ge 150\,mrad$, for the Bhabha process
$e^+e^-\rightarrow e^+e^-+n(\gamma)$.
\par
We turn first to the implementation of the required wide angle physics
effects in the YFS hard photon residual $\bar\beta_0$
through ${\cal O}(\alpha)$. We follow this discussion with the
corresponding analysis for the wide angle physics effects in the
hard photon YFS residual $\bar\beta_1$. In this way, 
we arrive at an exact ${\cal O}(\alpha)$ YFS exponentiated MC event
generator valid for wide angle Bhabha scattering in the LEP1/SLC and
LEP2 energy regimes.
    
%===================== SECTION 2 ======================================
\section{Born and ${\cal O}(\alpha)$ contributions to $\bar\beta_0$
for wide angle Bhabha scattering at high energies}

In this section we develop the Born and ${\cal O}(\alpha)$ contributions
to the YFS hard photon residual $\bar\beta_0$ for the wide angle Bhabha
scattering process. We use the low angle limit of the residual already
presented in ref.~\cite{bhlumi1:1989} as a starting point and as a limiting
case cross-check.\par
Specifically, the hard photon residual $\bar\beta_0$ is defined as
follows through ${\cal O}(\alpha)$:
\begin{equation}
\frac{1}{2}\bar\beta_0 = \frac{d\sigma^{1-loop}}{d\Omega} - 2\alpha\Re{B}
\,\frac{d\sigma_{Born}}{d\Omega}, 
\label{eq4}
\end{equation}
where $d\sigma^{1-loop}/d\Omega$ is the differential cross section
for wide angle Bhabha scattering computed through the 1-loop correction and
$d\sigma_{Born}/d\Omega$ is the respective Born differential cross
section. In ref.~\cite{bhlumi1:1989}, the right-hand side of (\ref{eq4}) was
computed in the low angle regime. Here, we need to compute (\ref{eq4})
in the wide angle scattering regime. We do this as follows. First, we 
note that the exact expression for the wide angle scattering
Born differential cross section
in~(\ref{eq4}) is well-known 
\begin{equation}
\frac{d\sigma_{Born}}{d\Omega}=\frac{1}{64\pi^2s}\,|{\cal M}_0|^2,
\label{dsig0}
\end{equation}
where $|{\cal M}_0|^2$ is a spin averaged lowest order matrix element squared, 
given e.g. in eq.~(\ref{mat0}). 
Secondly, the complete 1-loop corrected wide angle
Bhabha scattering differential cross section in (\ref{eq4}) is also known
and we use two different versions of it in our work, one taken from
ref.~\cite{bohm:1988} and one taken from ref.~\cite{alibaba:1991a}, where it
is generally accepted that the latter version, which is the more
recent of the two, is in fact the more up-to-date of the 
two~\cite{bhabha-lepws:1995} (and hence it 
is the option which we
illustrate in our comparisons in Sect.~4).
In our Monte Carlo event generator
these two versions for $d\sigma^{1-loop}/d\Omega$
correspond to two choices for our electroweak library module, one of
which the user specifies in his input file~\cite{bhwide-cpc}.
It is this way that we have realized the wide angle Bhabha
scattering YFS hard photon residual $\bar\beta_0$ through ${\cal O}(\alpha)$.
\par
We turn next to the wide angle Bhabha scattering YFS hard photon residual
$\bar\beta_1$. This we do in the following section.
   
%====================== SECTION 3 =====================================
\section{The YFS hard photon residual $\bar\beta_1$ for wide angle
Bhabha scattering at high energies}

In this section we present our determination of the ${\cal O}(\alpha)$
YFS hard photon residual $\bar\beta_1$ for wide angle Bhabha scattering
at high energies. We start with the defining equation for $\bar\beta_1$.
\par 
The hard photon residual $\bar\beta_1$, to ${\cal O}(\alpha)$, is defined
as follows
\begin{equation}
\frac{1}{2}\bar\beta_1 = \frac{d\sigma^{B1}}{kdkd\Omega_\gamma d\Omega}
 - \tilde S(k)\,\frac{d\sigma_{Born}}{d\Omega}, 
\label{eq5}
\end{equation}
where $d\sigma^{B1}/kdkd\Omega_\gamma d\Omega$
is the respective ${\cal O}(\alpha)$ bremsstrahlung differential cross 
section
into the solid angles $d\Omega_\gamma$ and $d\Omega$ for the photon
and positron respectively when the photon energy lies between $k$ and $k+dk$.
Thus, to specify $\bar\beta_1$ we need to specify the hard bremsstrahlung
differential cross section on the RHS of (\ref{eq5}). We now turn to this.
\par
Specifically, the hard bremsstrahlung differential cross section required
in (\ref{eq5}) can be obtained via the standard methods from the 
corresponding helicity amplitudes 
${\cal M}(\lambda_{e^+},\lambda_{e^-},$ $\lambda'_{e^+},\lambda'_{e^-},
\lambda_{\gamma})$, where $\lambda^{\tiny(')}_{f}$, $f=e^+,e^-$ 
is the incoming (outgoing) fermion helicity and $\lambda_{\gamma}$
is the photon helicity. We have
\begin{equation}
\frac{d\sigma^{B1}}{kdkd\Omega_{\gamma} d\Omega} = 
\frac{1}{512\pi^5 s}\, \frac{(p_2^0)^2}{2E_b(E_b-k)}
\,|{\cal M}|^2,
\label{eq6}
\end{equation}
with
\begin{equation}
|{\cal M}|^2 = \frac{1}{4}
\sum_{\underset{i=e^{\pm},\gamma,j=e^{\pm}}{\lambda_i,\lambda'_j}}
|{\cal M}(\lambda_{e^+},\lambda_{e^-},\lambda'_{e^+},\lambda'_{e^-},
\lambda_{\gamma})|^2
\label{mat1}
\end{equation}
being the spin averaged squared matrix element. The formula~(\ref{eq6})
is given in the CMS of the incoming beams, where $E_b$ is the beam energy.  
Thus, our determination of $\bar\beta_1$ will be 
completely specified when we give our results for 
${\cal M}(\lambda_{e^+},\lambda_{e^-},\lambda'_{e^+},\lambda'_{e^-},
\lambda_{\gamma})$. We now turn to this.
\par
More precisely, using the methods of ref.~\cite{xu:1987}, we get the
following representation for the non-vanishing helicity amplitudes: 
\begin{align}
M_1\:\equiv M(+++++)=-c\,s\:F_1  \bigg[\,&\frac{1}{t_p}R(t_p,a_La_R)\,G_1
                                        +\frac{1}{t_q}R(t_q,a_La_R)\,G_2  \,   
                                 \bigg]
\notag\\
M_2\:\equiv M(++++-)= -c\,s'F_2^{\ast}\bigg[\,&\frac{1}{t_p}R(t_p,a_La_R)\,
                                               G_1^{\ast}
                                              +\frac{1}{t_q}R(t_q,a_La_R)\,
                                               G_2^{\ast}\,
                                       \bigg]
\notag\\ 
M_3\:\equiv M(----+)= -c\,s'F_2  \bigg[\,&\frac{1}{t_p}R(t_p,a_La_R)\,G_1
                                       +\frac{1}{t_q}R(t_q,a_La_R)\,G_2  \,   
                                \bigg]
\notag\\
M_4\:\equiv M(-----)=-c\,s\:F_1^{\ast}\bigg[\,&\frac{1}{t_p}R(t_p,a_La_R)\,
                                               G_1^{\ast}
                                              +\frac{1}{t_q}R(t_q,a_La_R)\,
                                               G_2^{\ast}\,
                                      \bigg]
\notag\\ 
M_5\:\equiv M(-++-+)= -c\,t_q F_4  \bigg[\,&\frac{1}{s'}R(s',a_La_R)\,G_3
                                         +\frac{1}{s }R(s ,a_La_R)\,G_4  \, 
                                  \bigg]
\notag\\
M_6\:\equiv M(-++--)= -c\,t_p F_3^{\ast}\bigg[\,&\frac{1}{s'}R(s',a_La_R)\,
                                                 G_3^{\ast}
                                                +\frac{1}{s }R(s ,a_La_R)\,
                                                 G_4^{\ast}\,
                                        \bigg]
\notag\\ 
M_7\:\equiv M(+--++)= -c\,t_p F_3  \bigg[\,&\frac{1}{s'}R(s',a_La_R)\,G_3
                                         +\frac{1}{s }R(s ,a_La_R)\,G_4  \, 
                                  \bigg]
\notag\\
M_8\:\equiv M(+--+-)= -c\,t_q F_4^{\ast}\bigg[\,&\frac{1}{s'}R(s',a_La_R)\,
                                                 G_3^{\ast}
                                                +\frac{1}{s }R(s ,a_La_R)\,
                                                 G_4^{\ast}\,
                                        \bigg]
\notag\\ 
M_9 \: \equiv M(+-+-+)=+c\,u\:F_5  \bigg[\,&\frac{1}{t_p}R(t_p,a_L^2)\,G_1
                                          +\frac{1}{t_q}R(t_q,a_L^2)\,G_2
\notag\\
                                         +&\frac{1}{s' }R(s' ,a_L^2)\,G_3
                                          +\frac{1}{s  }R(s  ,a_L^2)\,G_4  \, 
                                   \bigg]
\notag\\
M_{10} \equiv M(+-+--)=+c\,u'F_6^{\ast}\bigg[\,&\frac{1}{t_p}R(t_p,a_L^2)\,
                                                G_1^{\ast}
                                               +\frac{1}{t_q}R(t_q,a_L^2)\,
                                                G_2^{\ast}
\notag\\
                                               +&\frac{1}{s' }R(s' ,a_L^2)\,
                                                 G_3^{\ast}
                                                +\frac{1}{s  }R(s  ,a_L^2)\,
                                                 G_4^{\ast}\, 
                                       \bigg]
\notag\\
M_{11} \equiv M(-+-++)=+c\,u'F_6  \bigg[\,&\frac{1}{t_p}R(t_p,a_R^2)\,G_1
                                          +\frac{1}{t_q}R(t_q,a_R^2)\,G_2
\notag\\
                                         +&\frac{1}{s' }R(s' ,a_R^2)\,G_3
                                          +\frac{1}{s  }R(s  ,a_R^2)\,G_4  \, 
                                   \bigg]
\notag\\
M_{12}\equiv M(-+-+-)=+c\,u\:F_5^{\ast}\bigg[\,&\frac{1}{t_p}R(t_p,a_R^2)\,
                                                G_1^{\ast}
                                               +\frac{1}{t_q}R(t_q,a_R^2)\,
                                                G_2^{\ast}
\notag\\
                                              +&\frac{1}{s' }R(s' ,a_R^2)\,
                                                G_3^{\ast}
                                              +\frac{1}{s  }R(s  ,a_R^2)\,
                                               G_4^{\ast}\, 
                                       \bigg]
\label{helamp}
\end{align}
where our notation is as follows
\begin{itemize}
\item Invariants:
\begin{align}
& s=2p_1q_1, \:\:\:\:\:\:\: s'=2p_2q_2,
\notag\\
& t_p=-2p_1p_2,\:\: t_q=-2q_1q_2,
\notag\\
& u=-2p_1q_2,\:\:\: u'=-2q_1p_2.
\label{Lorinv}
\end{align}

\item Electroweak couplings:
\begin{equation}
a_L=\frac{G}{e}\,(v_e + a_e), \:\:\: a_R=\frac{G}{e}\,(v_e - a_e),
\label{ewcc}
\end{equation}
where 
\begin{equation}
G=\frac{1}{\sin\theta_W\cos\theta_W},\:\:\: 
a_e=\frac{-1}{4\sin\theta_W\cos\theta_W},\:\: 
v_e=a_e(1-4\sin^2\theta_W).
\label{Gav}  
\end{equation}

\item Spinor functions:
\begin{align}
& F(p,q)=\frac{<pq>}{<pq>^{\ast}},
\notag\\ 
& G(p,q,r,s;k)=\frac{<pq>^{\ast}}{<rk><ks>},
\label{spifun}
\end{align}
where $p,q,r,s$ denote fermion four-momenta in the massless limit,
while $k$ denotes the 
photon four-momentum, and the ${}^\ast$ stands for complex
conjugation. Spinor products $<..>$ are defined as
\begin{equation}
<pq> = \sqrt{p_-q_+}\,e^{i\phi_p} - \sqrt{p_+q_-}\,e^{i\phi_q},
\label{spinpro}
\end{equation}
where $p_{\pm}=p_0 \pm p_z$, 
$p_{\bot}=p_x+ip_y=\sqrt{p_+p_-}\,e^{i\phi_p}$.

\item Propagator factor:
\begin{equation}
R(x,y) = 1 + \frac{xy}{x-M_{Z}^2+i\theta(x) M_{Z}\Gamma_{Z}}.
\label{propfa}
\end{equation}

\item Specific notation:
\begin{align}
& F_1=F(p_1,q_1),\:\: F_2=F(p_2,q_2),\:\: F_3=F(p_1,p_2),
\notag\\
& F_4=F(q_1,q_2),\:\: F_5=F(p_1,q_2),\:\: F_6=F(q_1,p_2),
\notag\\
& G_1=G(p_1,p_2,q_1,q_2;k),\:\: G_2=G(q_2,q_1,p_1,p_2;k),
\notag\\ 
& G_3=G(q_2,p_2,p_1,q_1;k),\:\: G_4=G(p_1,q_1,p_2,q_2;k),
\notag\\
&c=i2\sqrt{2}e^3.
\label{specnot}
\end{align}
\end{itemize}

The above helicity amplitudes have been obtained in the massless fermion 
approximation. However, to get a precise description of the photon
radiation over the whole phase space (particularly for collinear 
configurations) the finite fermion masses have to be taken into account.  
This can be accomplished by adding to the matrix element 
$|{\cal M}|^2$ of eq.~(\ref{mat1}) 
the mass correction term~\cite{calkul:1982}
\begin{align}
\delta|{\cal M}|_{mc}^2 =-e^2  \bigg[& 
  \frac{m_e^2}{(kp_1)^2} |{\cal M}_0(s',t_q,u')|^2 
 +\frac{m_e^2}{(kq_1)^2} |{\cal M}_0(s',t_p,u )|^2 
\notag\\
+&\frac{m_e^2}{(kp_2)^2} |{\cal M}_0(s ,t_q,u )|^2 
 +\frac{m_e^2}{(kq_2)^2} |{\cal M}_0(s ,t_p,u')|^2
                                \bigg],
\label{matmc}
\end{align}
where
\begin{align}
|{\cal M}_0(s,t,u)|^2 = e^4  \bigg\{ &
  \frac{1}{s^2}\big[ |R(s,a_L^2)|^2 u^2  + |R(s,a_R^2)|^2 u^2  
                    + 2 |R(s,a_La_R)|^2 t^2 \big]
\notag\\
+&\frac{1}{t^2}\big[ |R(t,a_L^2)|^2 u^2 + |R(t,a_R^2)|^2 u^2  
                    + 2 |R(t,a_La_R)|^2 s^2 \big]
\notag\\               
+&\frac{1}{st} \, 2\Re\big[
R^{\ast}(s,a_L^2) R(t,a_L^2) + R^{\ast}(s,a_R^2) R(t,a_R^2)
                      \big]    
                             \bigg\}
\label{mat0}
\end{align}
is the lowest order matrix element.  

We have also used the results of ref.~\cite{calkul:1982} for the matrix element
$|{\cal M}|^2$ as a cross check
and we have found that the two sets of results are in very good agreement
with one another, well below the desired technical precision of $0.01\%$ of
our analysis for example. We also note that an equivalent representation of the
results in ref.~\cite{calkul:1982} has been given by 
Kleiss in ref.~\cite{kleiss1:1987}. 
On introducing these two sets of results into the
formula (\ref{eq5}) for $\bar\beta_1$ and implementing the resulting
expression into our YFS Monte Carlo program for Bhabha scattering,
we arrive at the Monte Carlo event generator BHWIDE~1.00 for wide angle
Bhabha scattering at high energies\footnote{We need to stress that, in 
implementing the results (\ref{eq6})--(\ref{mat0}) and the corresponding
results from ref.~\cite{calkul:1982} into (\ref{eq5}), care must be 
taken to compute the two terms on the RHS of (\ref{eq6}) in the same
manner with the same massive (massless) limits of the corresponding
parts of each respective term so that the result for $\bar\beta_1$
is numerically stable; we have done this.}. 
\par
We will now illustrate the application of BHWIDE~1.00
to LEP1/SLC and LEP2 physics scenarios in the next section.

%===================== SECTION 4 ======================================
\section{Results and comparisons}

In this section, we present sample Monte Carlo data which we use
to compare our predictions from BHWIDE~1.00 to those of related 
calculations as reported in ref.~\cite{bhabha-lepws:1995}.
We discuss both LEP1/SLC energies and LEP2 energies. 

%======================= FIGURE 1 ======================
\begin{figure}[!ht]
\centering
%\htmlimage{scale=1.6}
%----------------------------
\setlength{\unitlength}{0.1mm}
\begin{picture}(1700,1700)
\put(0,0){\makebox(0,0)[lb]{
\epsfig{file=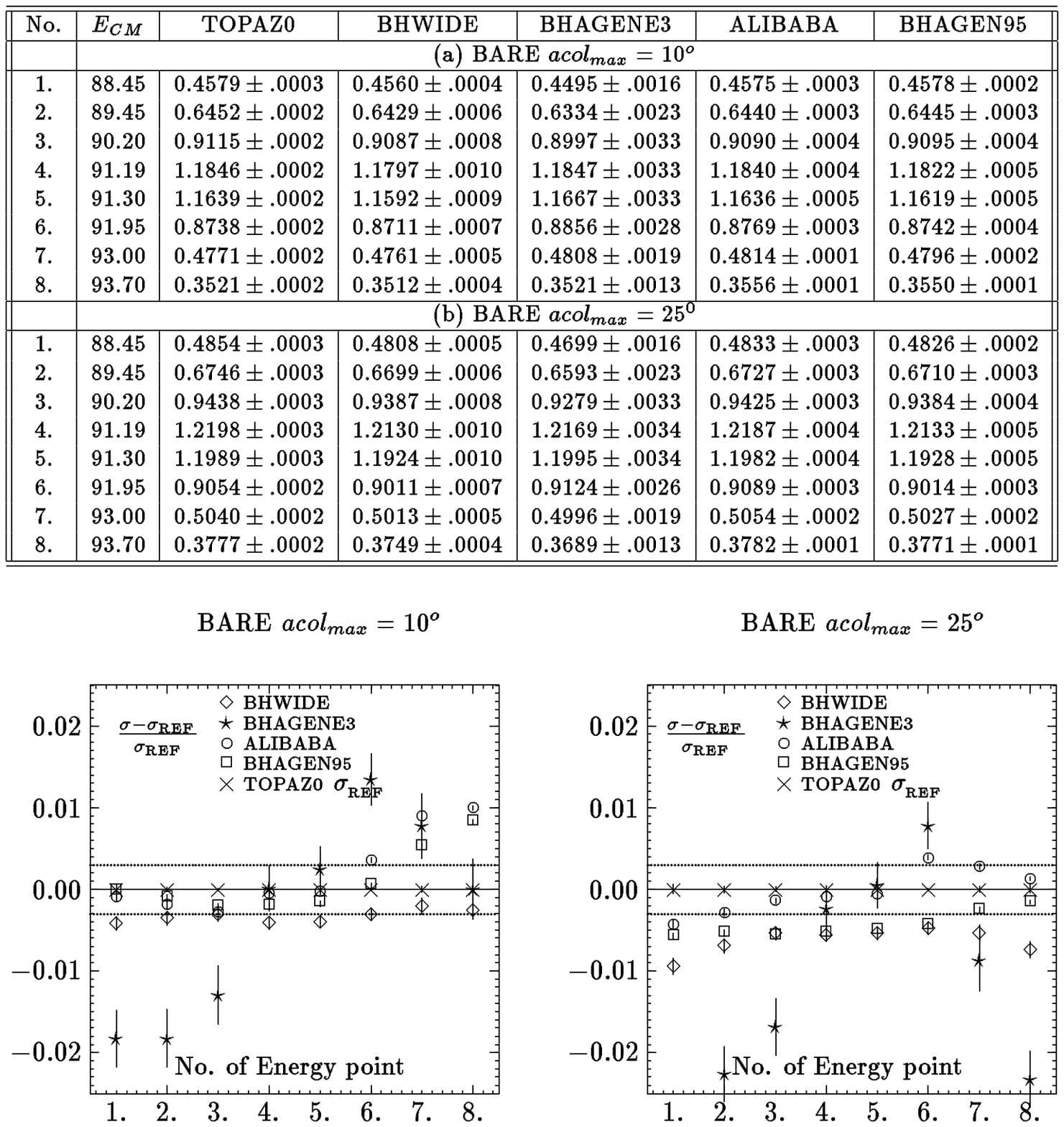,width=160mm,height=170mm} }}

\end{picture}

\caption{\small\sf
 Monte Carlo results for BARE trigger,
 for two values ($10^o$ and $25^o$) of acollinearity cut.
 Center of mass energies (in GeV) close to $Z$ peak.
 In the plots cross section
 $\sigma_{\rm{REF}}$ from TOPAZ0
 is used as a reference cross section. Cross sections in nb. 
 Two horizontal dotted lines indicate the $0.3\%$ band, for
 reference. 
% end-of-caption
}
\label{fig:lep1-bare}
\end{figure}
%================= END FIGURE 1 =======================

%======================= FIGURE 2 ======================
\begin{figure}[!ht]
\centering
%\htmlimage{scale=1.6}
%----------------------------
\setlength{\unitlength}{0.1mm}
\begin{picture}(1700,1700)
\put(0,0){\makebox(0,0)[lb]{
\epsfig{file=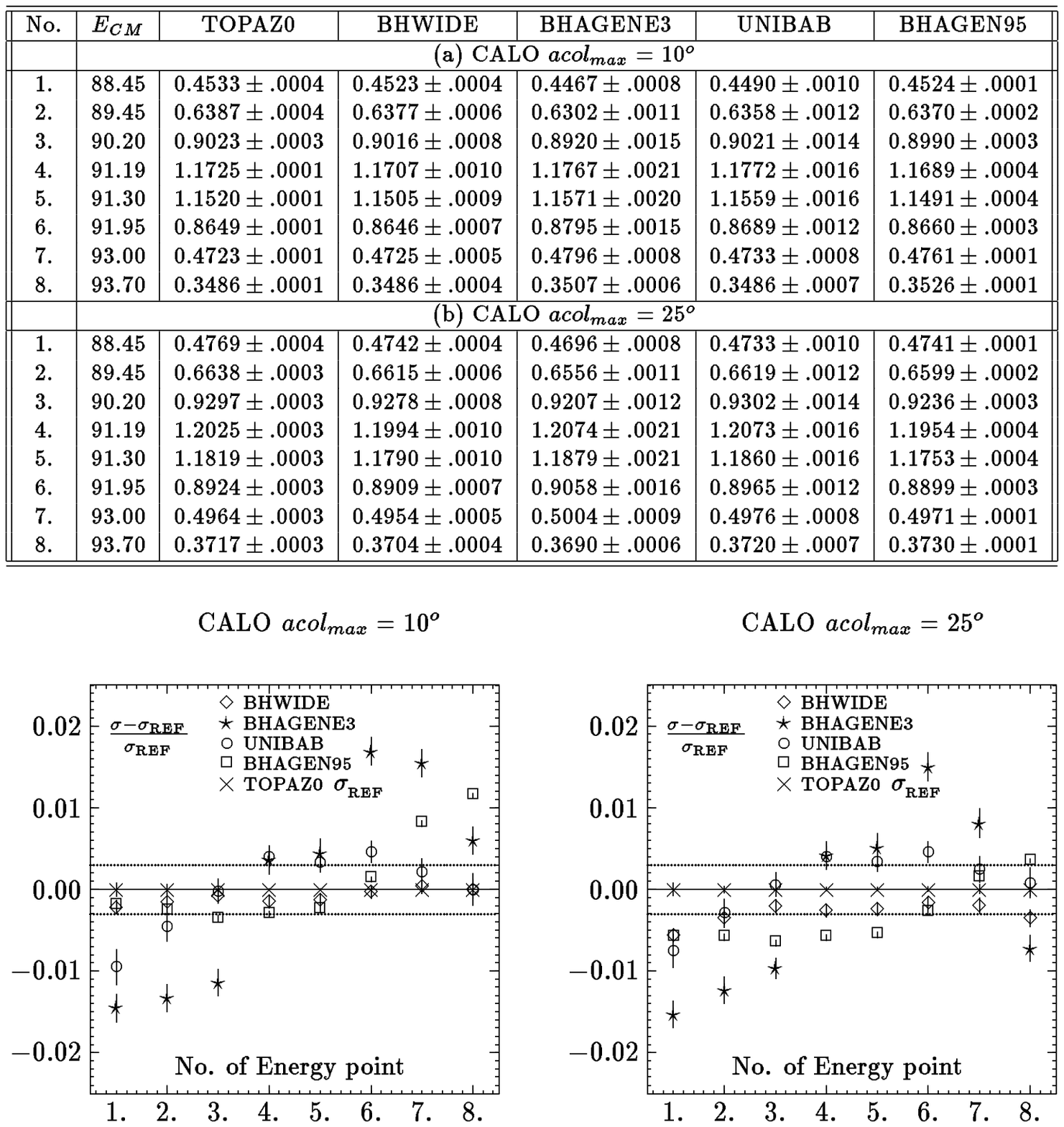,width=160mm,height=170mm} }}

\end{picture}
\caption{\small\sf
 Monte Carlo results for CALO trigger,
 for two values ($10^o$ and $25^o$) of acollinearity cut.
 Center of mass energies (in GeV) close to $Z$ peak.
 In the plots cross section
 $\sigma_{\rm{REF}}$ from TOPAZ0
 is used as a reference cross section. Cross sections in nb. 
 Two horizontal dotted lines indicate the $0.3\%$ band, for
 reference. 
% end-of-caption
}
\label{fig:lep1-calo}

\end{figure}
%================= END FIGURE 2 =======================

%======================= FIGURE 3 ======================
\begin{figure}[!ht]
\centering
%\htmlimage{scale=1.6}
%----------------------------
\setlength{\unitlength}{0.1mm}
\begin{picture}(1700,1300)
\put(0,0){\makebox(0,0)[lb]{
\epsfig{file=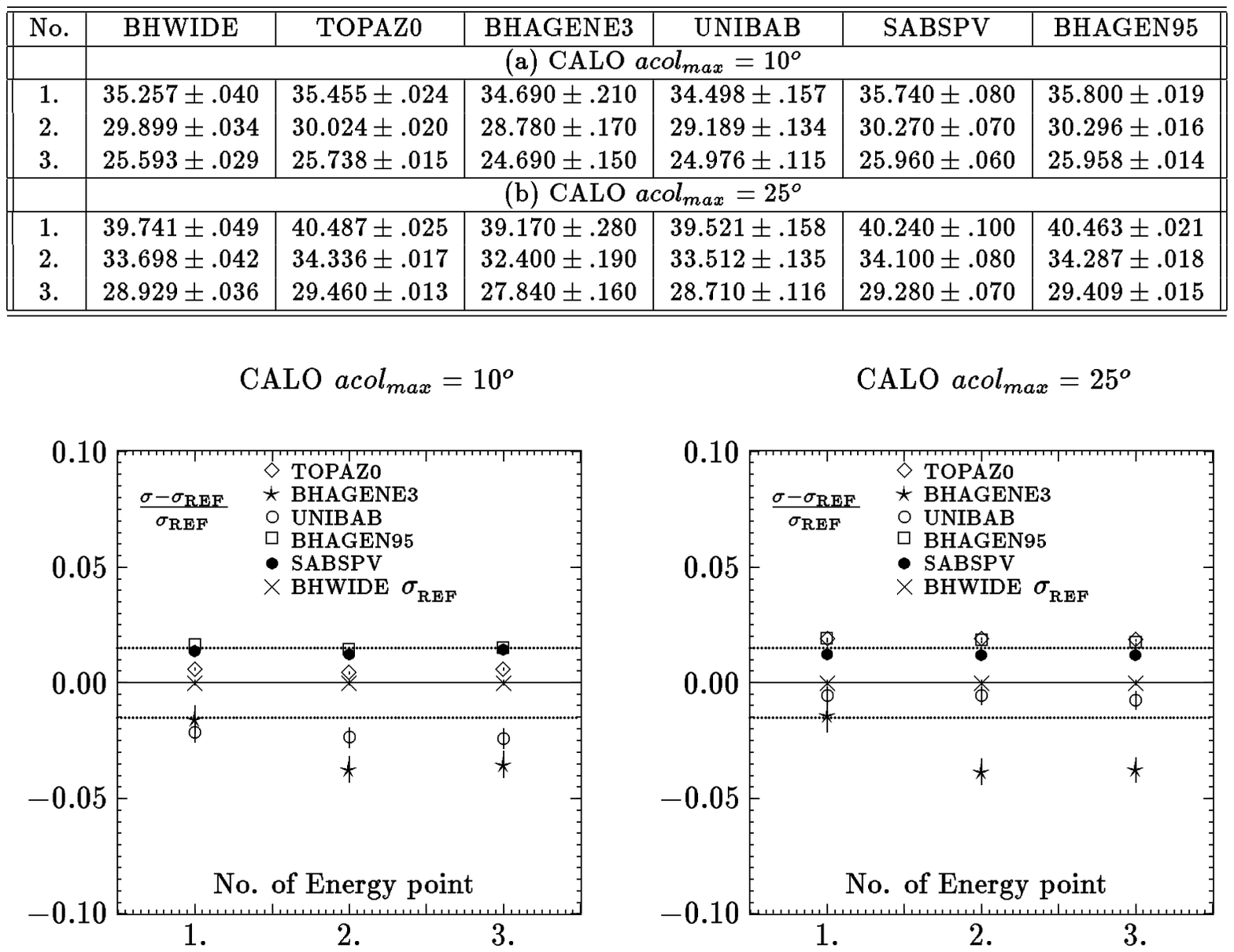,width=160mm,height=125mm} }}

\end{picture}
\caption{\small\sf
 Monte Carlo results for CALO trigger,
 for two values ($10^o$ and $25^o$) of acollinearity cut.
 Center of mass energies close to $W$-pair production threshold
 ($E_{CM}$: 1.~175~GeV, 2.~190~GeV, 3.~205~GeV).
 In the plots cross section
 $\sigma_{\rm{REF}}$ from BHWIDE
 is used as a reference cross section. Cross sections in pb. 
 Two horizontal dotted lines indicate the $1.5\%$ band, for
 reference. 
% end-of-caption
}
\label{fig:lep2-calo}
\end{figure}
%================= END FIGURE 3 =======================

\par
Specifically, for our comparisons we use the same event selection (ES) cuts
as defined in ref.~\cite{bhabha-lepws:1995}:
\begin{itemize}
\item{  BARE ES -- we require $40^o< \theta_{e^-}< 180^o$, 
$0^o< \theta_{e^+}< 180^o$, with acollinearity cuts of $10^o$, $25^o$ 
and $E_{min}= 1$ GeV for both $e^-$ and $e^+$;}
\item{  CALO ES -- we require the same cuts as in the BARE case but with 
$E_{min}= 20$ GeV for the final `fermion' energy which is the $e^-(e^+)$
energy if the there are no photons nearby and which is the $e^-(e^+)$
plus the photon energy if the photon is within a cone of half-angle
$1^o$ from the $e^-(e^+)$, respectively.}
\end{itemize}
For the BARE ES acceptance cuts, 
we show in fig.~\ref{fig:lep1-bare} the comparison of
the BHWIDE results with those of ALIBABA,
BHAGEN95, BHAGENE3 and TOPAZ0, where for definiteness
we plot the ratio $(\sigma_A-\sigma_{REF})/\sigma_{REF}$ for
each calculation $A$, $A=$ ALIBABA, BHAGEN95\footnote{
                          Here we use recently updated results of
                          BHAGEN95, obtained from the authors.}, 
BHAGENE3, BHWIDE, and TOPAZ0, 
using the TOPAZ0 cross section as $\sigma_{REF}$. This we do
for the 8 CMS energy values: $88.45$, $89.45$, $90.20$, $91.19$, $91.30$, 
$91.95$, $93.00$, $93.70\,GeV$, 
which are denoted in the figure as energy points $1,2,\ldots,8$,
respectively. 
We see that BHWIDE agrees with the semi-analytical programs, ALIBABA and
TOPAZ0, to within a few per mille at the $Z$ peak whereas off the
peak BHWIDE remains generally within $1\%$ of the semi-analytical programs.
The difference between BHWIDE and the Monte Carlo program BHAGENE3
is also at the few per mille level at the $Z$ peak but it is as much as
$2.27\%$ off the peak. The agreement between BHWIDE and BHAGEN95 is also
at the few per mille level at the $Z$ peak; off the peak, two programs
remain within $1\%$ of one another, with the better agreement holding
for the looser acollinearity cut of $25^o$.
These results give us a handle on the total precision of BHWIDE in the
$Z$ resonance regime, as we discuss presently.
\par
Continuing in this way, we show in fig.~\ref{fig:lep1-calo} the similar
type of comparison of the predictions of the programs with the CALO ES.
In fig.~\ref{fig:lep1-calo} ALIBABA is no longer applicable (it does not handle
the CALO ES) and UNIBAB appears --- it was too slow to
participate in the BARE ES comparisons.
Again, we use TOPAZ0 for the reference cross section and we plot the
same ratio $(\sigma_A-\sigma_{REF})/\sigma_{REF}$ for the same 8 energy
points as in fig.~\ref{fig:lep1-bare}. At the peak, BHWIDE is within a few
per mille of TOPAZ0 and BHAGEN95 and it is within 7 per mille of 
UNIBAB and BHAGENE3. Off peak, BHWIDE remains within 6 per mille of
TOPAZ0 whereas it remains within $0.8\%$ of UNIBAB, 
within $1.2\%$ of BHAGEN95 and 
within $1.8\%$ of BHAGENE3; the better agreement with the latter three
programs holds for the
looser acollinearity cut whereas for TOPAZ0 the situation is reversed.
Based on these and related comparisons as described in ref.~\cite{bhabha-lepws:1995}, including both the results and the physics approximations in
the various programs as presented in this last reference,
we conclude that, for the CALO ES,
within $\pm 100$ MeV of the $Z$ peak, the total precision of BHWIDE is 3
per mille and off peak, within $+2.5/-2.75$ GeV thereof, 
we set this precision at 5 per mille in the
LEP1 energy regime. For reference the three per mille band is indicated 
by the two horizontal dotted lines 
in figs.~\ref{fig:lep1-bare} and \ref{fig:lep1-calo}. This precision
tag should be compared to that for the only other published wide angle
Bhabha scattering Monte Carlo event generator in wide use at LEP/SLC,
namely BABAMC~\cite{babamc:1988}, whose total precision on pure QED was 
set at $1\%$ in the $Z$ peak region in refs.~\cite{lep1ybk:1989,fritsancy}.
(In practice, ALIBABA and/or TOPAZ0 would be used to determine the non-QED
and higher order QED part of the respective cross section.)
\par
Turning next to LEP2 energies, we show in fig.~\ref{fig:lep2-calo} the
comparison of the results of the six programs BHAGEN95, BHAGENE3, BHWIDE, 
SABSPV, TOPAZ0, and UNIBAB in the same format
as in figs.~\ref{fig:lep1-bare}, \ref{fig:lep1-calo}, 
using in this figure BHWIDE for the reference cross section $\sigma_{REF}$
for the three CMS energy points $175,\,190$, and $205\,GeV$. 
We use the CALO ES only here. SASPV and BHAGEN95 are within
$2\%$ of BHWIDE for both acollinearity cuts, with SASPV within
$1.5\%$ of BHWIDE and with BHAGEN95 maximally at $0.55\%$ above SASPV.
TOPAZ0 and UNIBAB deviate from BHWIDE by as much as
$1.9\%$ and $2.4\%$ respectively,
with the worse agreement holding for the looser (tighter) acollinearity 
cut, respectively.
These deviations are consistent with the leading logarithmic 
accuracy one expects for these two $Z$ peak optimized codes.
Similarly, the deviations between BHWIDE and BHAGENE3 by as much as $3.8\%$
as shown in fig.~\ref{fig:lep2-calo} are also consistent with what one
expects from the leading logarithmic $Z$ peak optimized nature of BHAGENE3. 
On the basis of these and related comparisons, we estimate the
total precision of BHWIDE at the LEP2 energies as $1.5\%$, 
conservatively. For reference, the $1.5\%$ band is indicated by the
two horizontal dotted lines in fig.~\ref{fig:lep2-calo}.
\par
We end this section by noting that, in addition to the comparisons just
presented, we have also checked that the pure QED ${\cal O}(\alpha)$
predictions of BHWIDE are within $0.05\%$ of those of the exact
${\cal O}(\alpha)$ MC OLDBIS of Ref.~\cite{bhlumi2:1992} --- a modernized
version of the MC of Ref.~\cite{oldbis:1983}. 
This gives us additional confidence in the technical component of 
our total precision estimate for BHWIDE~1.00.

%===================== CONCLUSIONS ====================================
\section{Conclusions}

In this paper, we have presented a new multiple photon Monte Carlo
for wide angle Bhabha scattering at LEP1/SLC and LEP2 energies
in which the respective multiple photon effects are realized on
an event-by-event basis and in which the infrared singularities
are cancelled to all orders in $\alpha$ via YFS exponentiation.
This Monte Carlo calculation contains the exact ${\cal O}(\alpha)$
result and features two choices for the respective pure weak
corrections at ${\cal O}(\alpha)$, that in ref.~\cite{alibaba:1991a}
and that in ref.~\cite{bohm:1988}. It thus corresponds to the 
exact ${\cal O}(\alpha)$ YFS exponentiated treatment of wide angle
Bhabha scattering.\par

We have illustrated our new calculation at both LEP1/SLC energies
and at LEP2 energies, for two types of event selection, the BARE
and CALO selections of ref.~\cite{bhabha-lepws:1995}, which feature two
choices of the acollinearity cut, $10^o$ and $25^o$. In our
illustrations, we compared our predictions with those of 
refs.~\cite{alibaba:1991a,sabspv:1995,topaz0:1995,bhagene3:1995,bhagen:1995,
unibab:1994}.
We found in general a good agreement of the various calculations, 
an agreement consistent with the levels of approximations
and realms of applicability of the respective codes.
In this way, we arrived at the precision tags of $0.3\%$ for
BHWIDE at the $Z$ peak and of $1.5\%$ at LEP2 energies.
The program is available from the authors at the WWW URL
http://enigma.phys.utk.edu/pub/BHWIDE/.
\par
In summary, we have developed a new Monte Carlo event generator
for wide angle Bhabha scattering at LEP1/SLC and LEP2 energies
in which the infrared singularities are cancelled to all orders
in $\alpha$ via YFS exponentiation of the respective multiple
photon effects. We look forward with excitement to its application
to LEP1/SLC and LEP2 data analyses.
\par

\vspace{0.5cm}
\noindent{\bf Acknowledgments}
\vspace{0.2cm}

\noindent
Two of us (S.~J. and B.F.L.~W.) thank Profs. G. Veneziano and
G. Altarelli for the support and kind hospitality of the CERN
Theory Division, where a part of this work was performed. One of
us (B.F.L. W.) thank Prof. C. Prescott of SLAC for the kind hospitality
of SLAC Group A while this work was completed.
 
%%%%%%%%%%%%%%%%%%%%%%%%%%%%%%%%%%%%%%%%%%%%%%%%%%%%%%%%%%%%%%%%%%%%%%%%%%%%
%%%%%%%%%%%%%%%%%%%%%%%%%%%%%%%%%%%%%%%%%%%%%%%%%%%%%%%%%%%%%%%%%%%%%%%%%%%%
%%%%%%%%%%%%%%%%%%%%%%%%%%%%%%%%%%%%%%%%%%%%%%%%%%%%%%%%%%%%%%%%%%%%%%%%%%%%
\bibliographystyle{prsty}
\bibliography{bhw}

\begin{thebibliography}{10}

\bibitem{alibaba:1991a}
W. Beenakker, F.~A. Berends, and S.~C. van~der Marck, Nucl. Phys. {\bf B349},
  323  (1991).

\bibitem{montagna:1993}
G. Montagna {\it et~al.}, Nucl. Phys. {\bf B401},  3  (1993).

\bibitem{yfs:1961}
D.~R. Yennie, S. Frautschi, and H. Suura, Ann. Phys. (NY) {\bf 13},  379
  (1961).

\bibitem{bhlumi1:1989}
S. Jadach and B.~F.~L. Ward, Phys. Rev. {\bf D40},  3582  (1989).

\bibitem{bohm:1988}
M. B{\"o}hm, A. Denner, and W. Hollik, Nucl. Phys. {\bf B304},  687  (1988),
  and references therein.

\bibitem{babamc:1988}
F.~A. Berends, R. Kleiss, and W. Hollik, Nucl. Phys. {\bf B304},  712  (1988).

\bibitem{calkul:1982}
F.~A. Berends {\it et~al.}, Nucl. Phys. {\bf B206},  61  (1982).

\bibitem{kleiss:1985}
R. Kleiss and W.~J. Stirling, Nucl. Phys. {\bf B262},  235  (1985).

\bibitem{xu:1987}
{Zhan Xu}, {Da-Hua Zhang}, and {Lee Chang}, Nucl. Phys. {\bf B291},  392
  (1987).

\bibitem{sabspv:1995}
M. Cacciari, G. Montagna, O. Nicrosini, and F. Piccinini, Comput. Phys. Commun.
  {\bf 90},  301  (1995).

\bibitem{topaz0:1993}
G. Montagna {\it et~al.}, Comput. Phys. Commun. {\bf 76},  328  (1993).

\bibitem{topaz0:1995}
G. Montagna, O. Nicrosini, G. Passarino, and F. Piccinini, \uppercase{CERN}
  preprint CERN-TH.7463/94, June 1995, submitted to Comput. Phys. Commun.
  (unpublished).

\bibitem{jfield:1994}
J.~H. Field, Phys. Lett. {\bf B323},  432  (1994).

\bibitem{bhagene3:1995}
J.~H. Field and T. Riemann, report UGVA-DPNC 1995/6-166, DESY 95-100, to be
  published in Comput. Phys. Commun. (unpublished).

\bibitem{bhagen:1992}
M. Caffo, H. Czy{\.z}, and E. Remiddi, Nuovo Cim. {\bf 105A},  277  (1992).

\bibitem{bhagen:1993}
M. Caffo, H. Czy{\.z}, and E. Remiddi, Int. J. Mod. Phys. {\bf 4},  591
  (1993).

\bibitem{bhagen:1994}
M. Caffo, H. Czy{\.z}, and E. Remiddi, Phys. Lett. {\bf B327},  369  (1994).

\bibitem{bhagen:1995}
M. Caffo, H. Czy{\.z}, and E. Remiddi, {\em \uppercase{B}HAGEN95}, in
  preparation (unpublished).

\bibitem{unibab:1994}
H. Anlauf {\it et~al.}, Comput. Phys. Commun. {\bf 79},  466  (1994).

\bibitem{bhabha-lepws:1995}
S. Jadach {\it et~al.},  in {\em Physics at LEP2}, edited by G. Altarelli, T.
  Sj{\"o}strand, and F. Zwirner (CERN, Geneva, 1996), Vol.~2, p.\ 229,
  \uppercase{Y}ellow \uppercase{R}eport \uppercase{CERN} 96-01, e-Print
  Archive: hep-ph/9602393.

\bibitem{sjbw:1988}
S. Jadach and B.~F.~L. Ward, Phys. Rev. {\bf D38},  2897  (1988).

\bibitem{bhlumi2:1992}
S. Jadach, E. Richter-W\c{a}s, B.~F.~L. Ward, and Z. W\c{a}s, Comput. Phys.
  Commun. {\bf 70},  305  (1992).

\bibitem{bennie:1987}
B.~F.~L. Ward, Phys. Rev. {\bf D36},  939  (1987), \underline{ibid.} {\bf 42}
  3249 (1990).

\bibitem{yfs2:1990}
S. Jadach and B.~F.~L. Ward, Comput. Phys. Commun. {\bf 56},  351  (1990).

\bibitem{yfs3-unpub}
S. Jadach and B.~F.~L. Ward, Phys. Lett. {\bf B274},  470  (1992).

\bibitem{bhwide-cpc}
S. Jadach, W. P\l{}aczek, and B.~F.~L. Ward, {\em \uppercase{B}HWIDE 1.00, Long
  Write-Up}, in preparation, to be submitted to Comput. Phys. Commun.
  (unpublished).

\bibitem{kleiss1:1987}
R. Kleiss, Z. Phys. {\bf C33},  433  (1987).

\bibitem{lep1ybk:1989}
R. Kleiss {\it et~al.},  in {\em Z PHYSICS AT LEP 1}, edited by G. Altarelli,
  R. Kleiss, and C. Verzegnassi (CERN, Geneva, 1989), Vol.~3, pp.\ 1--142,
  \uppercase{Y}ellow \uppercase{R}eport \uppercase{CERN} 89-08.

\bibitem{fritsancy}
F. Berends,  in {\em ANNECY MEETING ON Z PHYSICS AT LEP 1}, edited by C.
  Verzegnassi (LAPP-Annecy, Annecy, 1990), transparencies.

\bibitem{oldbis:1983}
F.~A. Berends and R. Kleiss, Nucl. Phys. {\bf B228},  537  (1983).

\end{thebibliography}
%%%%%%%%%%%%%%%%%%%%%%%%%%%%%%%%%%%%%%%%%%%%%%%%%%%%%%%%%%%%%%%%%%%%%%%%%%%%%
%%%%%%%%%%%%%%%%%%%%%%%%%%%%%%%%%%%%%%%%%%%%%%%%%%%%%%%%%%%%%%%%%%%%%%%%%%%%
%%%%%%%%%%%%%%%%%%%%%%%%%%%%%%%%%%%%%%%%%%%%%%%%%%%%%%%%%%%%%%%%%%%%%%%%%%%

\end{document}